\newcommand\GeV{{\rm ~GeV}}

\newcommand{\Pom}{{I\!\!P}}
\newcommand{\xP}{x_{\!\Pom}}
\documentclass[epj]{svjour}
\usepackage{graphics}
\begin{document}
\title{Factorisation breaking in diffractive dijet photoproduction at HERA?}
\author{Radek \v Zleb\v c\'ik, Karel \v Cern\'y, Alice Valk\'arov\'a
}                     
%
%
\institute{Institute of Particle and Nuclear Physics of the Faculty of Mathematics 
        and Physics of Charles University, 
        V Hole\v sovi\v ck\'ach 2, 180 00 Praha 8, Czech Republic}
\date{Received: date / Revised version: date}
%
\abstract{
Recent experimental data on dijet cross sections in diffractive photoproduction  at HERA collider are analysed with an emphasis on QCD factorisation breaking effects. The possible sources of the different conclusions of H1 and ZEUS collaborations are studied.
\PACS{
      {PACS-key}{discribing text of that key}   \and
      {PACS-key}{discribing text of that key}
     } 
} 
\maketitle
\section{Introduction}
\label{intro}
The first observation of deep inelastic scattering (DIS) events at HERA containing a large gap in the pseudorapidity distributions of final hadrons \cite{FirstDiffraction,SecDiffraction} has generated a considerable renewed interest in understanding of diffraction. In the $ep$ diffractive interactions measured at the HERA experiments the proton can stay intact or dissociate into a low mass state (Y), while the photon dissociates into a hadronic state X, $\gamma^* p \to Xp' (Y)$.  The systems are separated by a non-exponentially suppressed large rapidity gap being present due to vacuum quantum numbers of the diffractive exchange \cite{Bjorken1}. The diffractive exchange, called pomeron ($\Pom$), carries away a fraction $\xP$ of the initial proton longitudinal momentum and a four-momentum transfer $t=(P-P_{Y})^2$ associated with the outgoing (leading) system Y \footnote{ where $P$ is the four-momentum of incoming proton}.

One of the most interesting questions which are discussed in the studies of diffractive processes is whether they can be considered being factorisable into diffractive parton distribution functions (DPDFs) of the proton and perturbatively calculable partonic cross sections. Such a concept, called hard QCD (or collinear) factorisation, was proved to be valid in the regime of diffractive deep inelastic scattering (DDIS) \cite{Collins}. On the other hand the factorisation concept was found not to be valid for hard processes in diffractive hadron-hadron scattering as measured for example in $p\bar{p}$ interactions at Tevatron \cite{Tevatron}. The predictions of cross sections of diffractive dijet production in $p\bar{p}$ based on DPDFs provided by the H1 collaboration, overestimate the measured cross section almost by one order of magnitude. This observation was explained later on theoretically \cite {Kaidalov} assuming that factorisation breaking results from absorptive effects caused by multiple rescattering effects, see e.g.  ~\cite{Dokshitzer,Bjorken} for earlier discussion. The rapidity gap can then be populated by secondary particles  which spoil the experimental signature of the diffractive event. 
 
At HERA, diffraction has been studied in a wide range of exchanged photon virtualities, $Q^2$, ranging from photoproduction regime, $Q^2\sim 0$, to the deep inelastic scattering with $Q^2\gg 0$. Various hard diffractive processes were studied at HERA such as inclusive ones, production of jets, charm and vector mesons.

The factorisation in diffractive DIS dijet production was experimentally tested by the H1 and ZEUS collaborations \cite{H1_DIS,ZEUS_DIS}. Far from clear, however, is the situation in the photoproduction regime of the $ep$ diffractive scattering. In photoproduction, in the leading order approximation, the small photon virtuality allows for partonic fluctuations that live long enough. The photon may not couple directly to the quarks in the proton, but only a part of its four-momentum participates in the hard interaction. Such interactions are called resolved. The photon can still couple directly (with its whole four-momentum) to the quarks and these interactions are called direct. The resolved photon interactions resemble the hadron-hadron ones since two particles with structure scatter on each other. The variable $x_{\gamma}$, which is defined as a four-momentum fraction of the photon taking part in the hard interaction, is used to distinguish between the two regimes in photoproduction. Obviously, following relations hold: $x_{\gamma}=1$ and $x_{\gamma}<1$ for the direct and resolved photon interactions, respectively. Effects of fragmentation and a finite experimental resolution impose a smearing on the value of $x_{\gamma}$. In experiments, often a value of $x_{\gamma}$ around $0.75$ is considered to be a discriminator below (above) which the events are regarded as being due to resolved (direct) photon interaction, with reasonably low contamination of one in each other. 

Because of the presence of these two photon interaction regimes, studying diffractive processes in photoproduction is a useful tool for tests of the validity of the hard factorisation in diffraction. The partonic structure of the photon is described in terms of quark and gluon densities that obey DGLAP evolution equations \cite{DGLAP}. There was an earlier prediction of Kaidalov et al.~\cite{Khoze} that the factorisation breaking of the resolved part should induce a suppression of the NLO QCD expectation by about  a factor of  $0.34$. This idea was widely discussed and applied to published data in the studies of Klasen and Kramer \cite{Klasen,Klasen_new}. However recently \cite{Khoze_new} theoretical expectations were revised stressing the fact that due to the inhomogeneous term in the DGLAP evolution there is also point-like part of the photon structure function ~\cite{Witten}. The hadron-like part of the photon structure (suppressed by 0.34) occurs only at lowest values of four-momentum fractions $x_{\gamma}\sim 0.1$ which are experimentally hardly accessible. The dominant measured part of the resolved processes is therefore induced by the point-like component of the photon structure function with a significantly weaker suppression as compared to the $0.34$, see ~\cite{Khoze_new} for details.

\begin{table}
\begin{center}
\caption{Kinematic cuts used in H1 \cite{Karel_new} and ZEUS \cite{Z_dijetsPH} ~~~~~~~~~~analyses}
\label{tab:1}       
\begin{tabular}{ll}
\hline\noalign{\smallskip}
H1 & ZEUS \\
\noalign{\smallskip}\hline\noalign{\smallskip}
$Q^2 < 0.01\, \mathrm{GeV^2}$ & $Q^2 < 1\, \mathrm{GeV^2}$\\
$0.3 < y < 0.65$ & $0.2 < y < 0.85$\\
$E_T^{\mathrm{jet}1} > 5\, \mathrm{GeV}$ & $E_T^{\mathrm{jet}1} > 7.5\, \mathrm{GeV}$\\
$E_T^{\mathrm{jet}2} > 4\, \mathrm{GeV}$ & $E_T^{\mathrm{jet}2} > 6.5\, \mathrm{GeV}$\\
$-1 < \eta^{\mathrm{jet1(2)}} < 2 $ & $-1.5 < \eta^{\mathrm{jet1(2)}} < 1.5 $\\
\hline
Diffractive cuts\\
\hline
$x_{I\!\!P} < 0.03$ & $x_{I\!\!P} < 0.025$\\
$z_{I\!\!P} < 0.8$ &\\
$|t| < 1 \, \mathrm{GeV^2}$ & $|t| < 1 \, \mathrm{GeV^2}$\\
$M_Y < 1.6\, \mathrm{GeV}$ & $M_Y = m_p$\\
\noalign{\smallskip}\hline
\end{tabular}
\end{center}
\end{table} 
 

\section{Recent results from HERA}

\label{sec:1}

The H1 and ZEUS  collaborations analysed the diffractive dijet photoproduction data in ~\cite{Sebastian,Z_dijetsPH,ZEUS_QCD}.
The advantage of using the diffractive photoproduction of dijets is that two jets in the final state enable us to reconstruct the $x_{\gamma}$ variable. The H1 collaboration 
 observed a global suppression of the dijet cross sections with respect to NLO QCD calculations by a factor of about 0.5 ~\cite{Sebastian}.
On the contrary
 the ZEUS data published in \cite{Z_dijetsPH,ZEUS_QCD} were compatible with no suppression.  A new study of H1 collaboration \cite{Karel_new} with a three times larger data sample and a similar kinematic region as the previous study ~\cite{Sebastian} fully confirms the previous H1 observation. 
In contradiction with expectations of only the resolved processes being suppressed \cite{Khoze,Khoze_new} neither experiment did observe a difference in suppression for the resolved enriched part ($x_\gamma < 0.75$) and the direct enriched part of the cross section ($x_\gamma > 0.75$).

\begin{figure}
\resizebox{0.48\textwidth}{!}{%
  \includegraphics{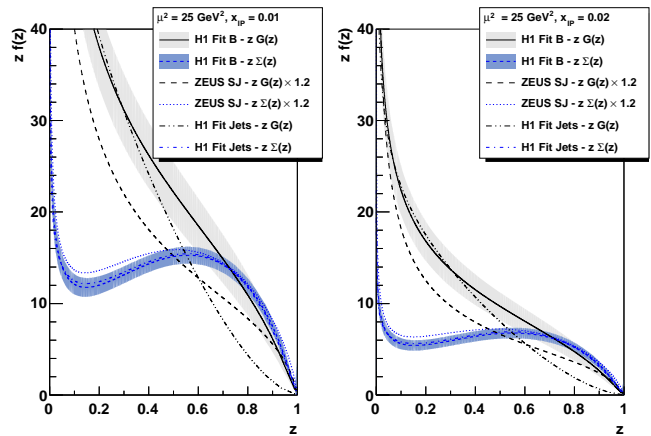}
}
\caption{The diffractive quark singlet and gluon densities for the squared factorisation scale $\mu^2=25\GeV^2$
in the region $M_Y < 1.6 \GeV$ and $|t| < 1 \GeV^2$ for two values of $x_\Pom$, $x_\Pom = 0.01$ (left) and $x_\Pom = 0.02$ (right). Uncertainties of the H1 fit B are depicted as a band. }
\label{fig:DPDFs}
\end{figure}

\begin{figure*}
\begin{center}
\resizebox{0.9\textwidth}{!}{%
  \includegraphics{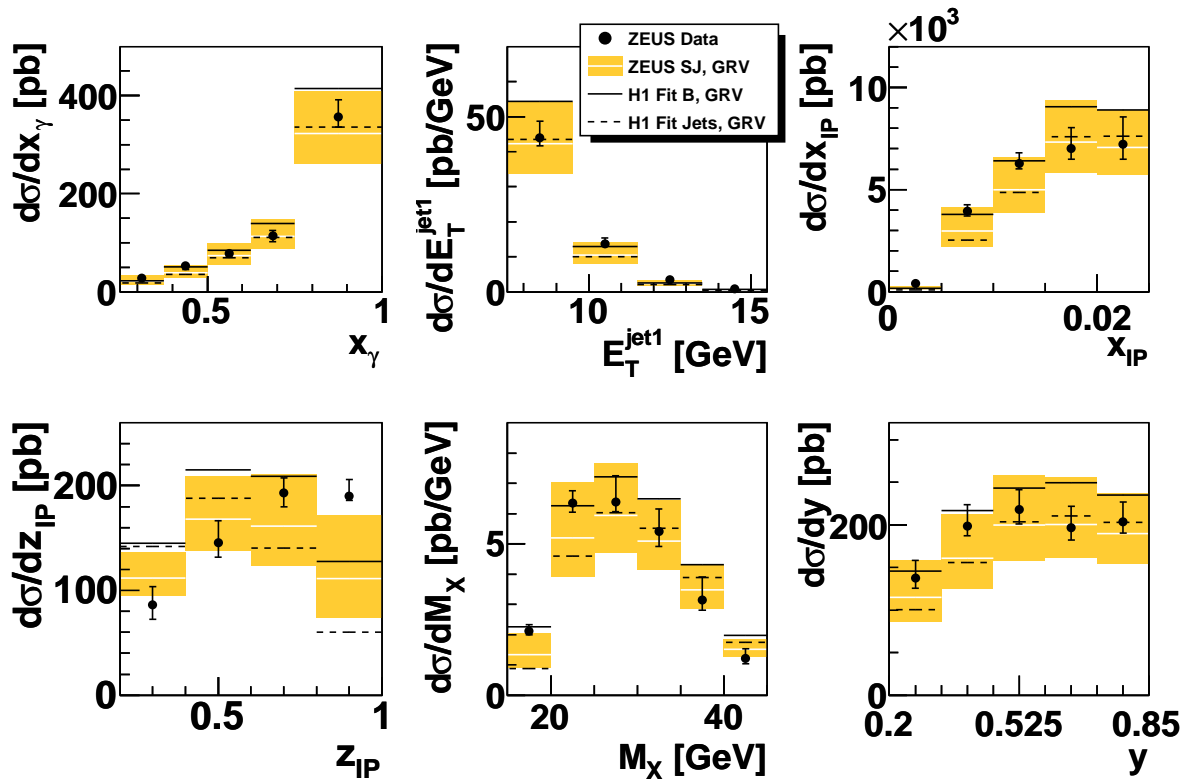}
}
\end{center}
\caption{Differential cross sections for diffractive dijet photoproduction as published by ZEUS \cite{Z_dijetsPH} 
compared with NLO QCD calculations corrected for hadronisation using DPDF ZEUS fit SJ (white line), and two H1 DPDFs divided by 1.2 - H1 fit B (full line) and H1 fit Jets (dashed line) . For NLO QCD calculations ZEUS fit SJ, the theoretical uncertainties connected with
the change of scale are shown as a dark band. }
\label{fig:3}       
\end{figure*}

In H1 and ZEUS analyses the diffractive events were selected with a large rapidity gap method and jets were identified using the inclusive $k_{\rm T}$ cluster algorithm \cite{kt} in the laboratory frame.
The phase space of both analyses was different (see Table 1), the main difference being at somewhat larger transverse momenta of the jets,  $E_{\rm T}^{\mathrm{jet}}$, of ZEUS measurement. H1 collaboration collected data with a tagged electron that allowed to restrict the $Q^2$ to very low values ($Q^2 < 0.01 \GeV^2$), the ZEUS analysis was done with an untagged electron sample of events ($Q^2 < 1 \GeV^2$). In H1 analysis the additional cut $z_\Pom < 0.8$ was applied since the DPDF sets are not valid at the largest values of $z_\Pom$.\footnote{ $z_\Pom$ is the longitudinal four-momentum fraction of the parton entering the hard subprocess with respect to pomeron.}

The NLO QCD calculations were performed by means of using the program of Frixione et al. \cite{Frixione} (H1) and Klasen and Kramer~\cite{KK} (ZEUS). Both calculations provide a consistent results as it was demonstrated in \cite{Karel}. 

For the NLO QCD calculations, the H1 collaboration used three DPDF sets, H1 2006 DPDF fit B, measured in the analysis of the inclusive DIS data \cite{H1_incl}, the inclusive and dijets combined fit H1 2007 DPDF fit Jets \cite{H1_DIS}  and inclusive and dijets   combined ZEUS fit SJ \cite{ZEUS_QCD}. 
H1 and ZEUS collaborations compared their results  with the calculations using the same DPDF fits.

 In Fig. 1. the comparison of the three DPDF sets is shown as a function of the partonic longitudinal four-momentum
 fraction with respect to the pomeron, $z$, for two values of $x_\Pom$.
 As expected a good agreement of all three fits is seen for the quark singlet density. The differences in the fits are much larger for gluon densities which are  more important for dijet measurements.  Note that the values of ZEUS fit SJ in the Fig. 1 are multiplied by a factor of 1.2. This is connected with the difference in methods which were used to measure diffractive inclusive cross section by H1 and ZEUS collaborations. The selection of diffractive events relying on  the rapidity gap method yields a sample which is dominated by elastically scattered protons, but which also contains an admixture of events in which the proton dissociates to low mass ($M_Y$) state. In the case of H1 collaboration the measurements are corrected to the region 
 $M_Y < 1.6 \GeV$ and $|t| < 1 \GeV^2$ while ZEUS collaboration corrects the measurements to $M_Y = m_p$, (for more details see e.g. \cite{H1_incl}).
 
\section{Crosscheck of H1 and ZEUS results}
 In the following the crosscheck and discussion of the possible sources for different H1 and ZEUS conclusions as well as the sensitivity to an alternative photon parton distribution function ($\gamma$-PDF) choice will be studied.

\begin{figure*}
\begin{center}
\resizebox{0.9\textwidth}{!}{%
  \includegraphics{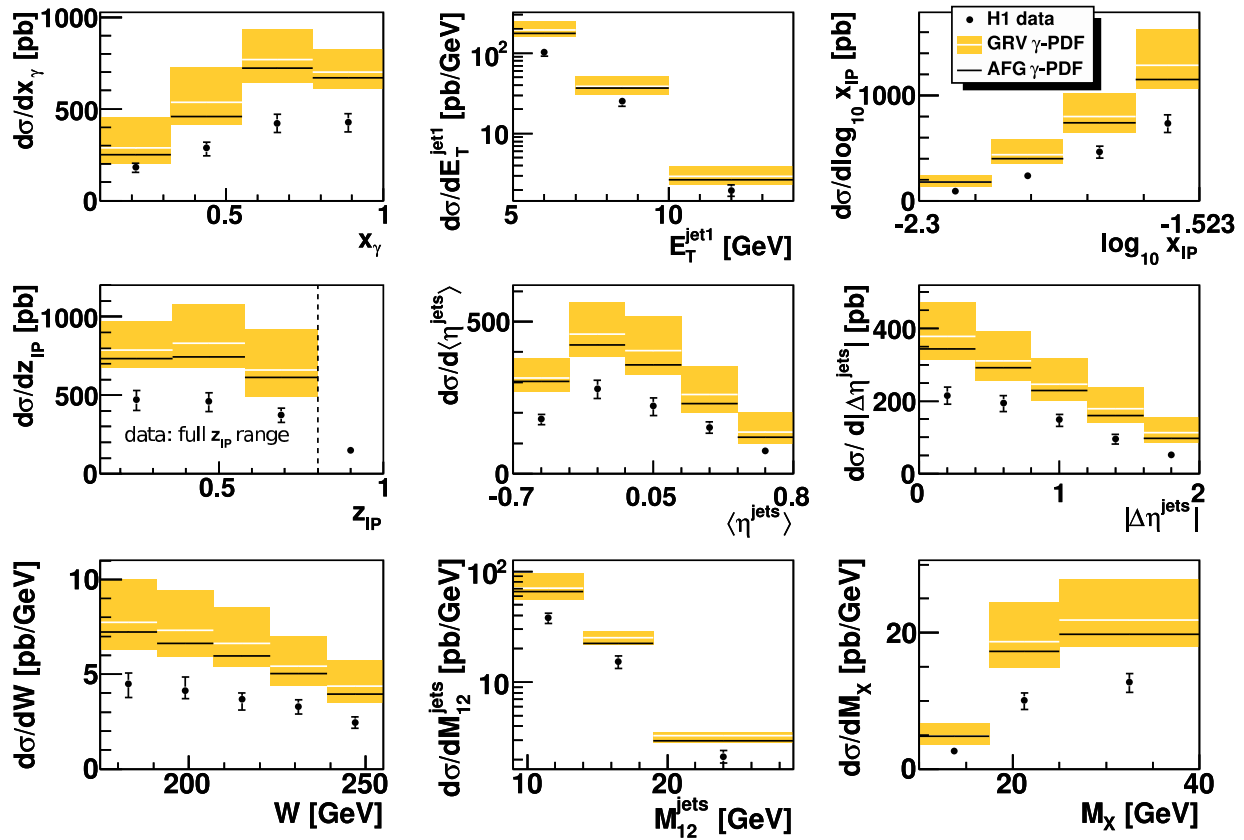}
}
\end{center}
\caption{The differential cross sections for diffractive dijet photoproduction as measured by H1 \cite{Karel_new}
compared with NLO calculations using H1 fit B DPDF and photon structure functions GRV and AFG. The dark bands
around the theoretical GRV PDF predictions (marked by white line) show the uncertainty of NLO calculations connected with the scale variation by factors 0.5 and 2.0.
 }
\label{fig:4}       
\end{figure*}

\begin{figure*}
\begin{center}
\resizebox{0.9\textwidth}{!}{%
  \includegraphics{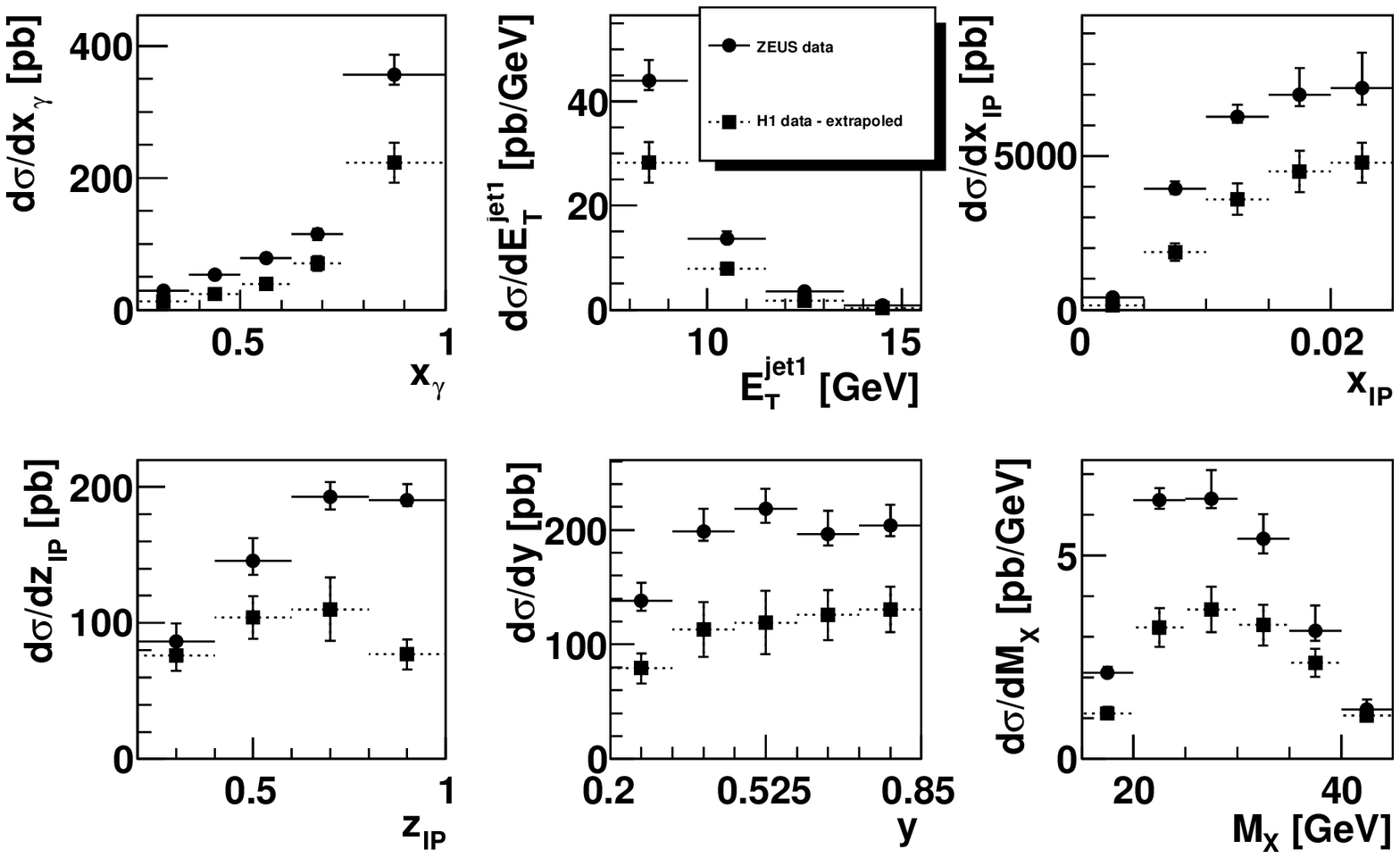}
}
\end{center}
\caption{Differential cross sections for diffractive dijet photoproduction as published by ZEUS \cite{Z_dijetsPH}      (points and full line)
compared to extrapolated H1 differential cross section (squares and dotted line).}
\label{fig:5}       
\end{figure*}

\paragraph{Hadronisation corrections  -}
To be able to compare the NLO QCD cross sections calculated at the level of partons with the data, the correction for hadronisation effects must be done. A common practice is to use Monte Carlo leading order models with initial and final state  parton radiation in order to mimic some features of the next-to-leading order parton dynamics. In \cite{Khoze_new} the justification of this procedure is discussed. It is noticed that the hadronisation corrections based on the LO Monte Carlo should be different from the effects associated with NLO calculations. To minimize this effect both collaborations used the procedure of reweighting of MC parton level distributions to NLO calculated parton level distributions. Such a procedure ensures that the bin-to-bin migrations due to the transitions from partons to hadrons will be described more accurately. The hadronisation corrections calculated and applied to NLO predictions were however different in H1 and ZEUS analyses. The hadronisation corrections, $\delta_{hadr}$, (defined in each bin $i$ as $(1+\delta_{hadr})_{i} \equiv C_i= \sigma_i^{hadr}/\sigma_i^{part}$, where $C_{i}$ is the correction factor to be applied to the NLO prediction) obtained in \cite{Sebastian} and independently also in \cite{Karel_new} are on average at the level of $-15$\%.  The hadronisation corrections calculated in \cite {Z_dijetsPH} and applied in \cite{Z_dijetsPH,ZEUS_QCD} are also at the level of 15\% but positive.  Nevertheless, it is evident that the difference in hadronisation corrections cannot explain the contradiction in conlusions of the two analyses. Identical hadronisation corrections would eventually make the difference in suppression even larger.
\vspace*{-0.2cm}

 An attempt was made to recalculate the hadronisation corrections in the kinematics of the analyses of both H1 and ZEUS experiments. 
  The recalculated hadronisation corrections for H1 kinematics agree within few percent with the published \cite{Karel_new} ones while 
 the recalculated hadronisation correction factors for ZEUS kinematics are lower by about 15\% than those from \cite {Z_dijetsPH}.
 \footnote{Note that hadronisation corrections in ZEUS case were evaluated in \cite {Z_dijetsPH} using the older DPDF than DPDF 
 ZEUS fit SJ used here.} In Fig. 2, the measured ZEUS cross sections for variables $x_\gamma$, $E_{\rm T}^{\mathrm{jet1}}$, $ x_\Pom$, $z_\Pom$, invariant mass of diffractive hadronic system $M_X$ and inelasticity $y$ from~\cite{Z_dijetsPH} are compared with the NLO predictions corrected by means of using of the recalculated hadronisation corrections. The experimental differential cross sections are clearly in agreement with NLO calculations using all DPDFs except for $d\sigma/dz_\Pom$ where the shapes of experimental and predicted distributions disagree. As expected the conclusions of both analyses as concerns the non-observation or observation of the factorisation breaking are not sensitive to hadronisation corrections used.

\paragraph{Alternative photon distribution function -}
In the analyses mentioned above the photon GRV structure function \cite{GRV} was used in the NLO calculations. It was noticed \cite{Khoze_new} that the point-like part of the resolved contribution in GRV $\gamma$-PDF may be overestimated by about 25\% in comparison with the more recent AFG $\gamma$-PDF \cite{AFG}.   In Fig.~3 the H1 experimental differential cross sections \cite{Karel_new} are shown for variables: $x_\gamma$, $E_{\rm T}^{\mathrm{jet1}}$, $\log_{10} x_\Pom$, $z_\Pom$, mean pseudorapidity of two jets, $\langle\eta^{\mathrm{jets}}\rangle$, absolute value of the difference in pseudorapidities of both jets,
$|\Delta\eta^{\mathrm{jets}}|$, total hadronic energy $W$, the invariant mass of jets $M_{12}$ and $M_X$. These experimental distributions are compared with NLO QCD predictions based on GRV and alternatively the AFG $\gamma$-PDF corrected for the effects of hadronisation. Indeed, it can be seen that the NLO QCD calculation based on AFG $\gamma$-PDF predicts smaller cross section than the one provided by GRV $\gamma$-PDF. In any case the difference in the value of suppression is not significant and is covered by the experimental uncertainties and by the variation of the renormalization and factorisation scale ($\mu_{r,f}/2$, $2\mu_{r,f}$). We can summarize that the conclusions of both HERA experiments are not sensitive to the $\gamma$-PDF used.

\paragraph{Matching the different phase space -}
The H1 and ZEUS dijet data cannot be compared directly, since they have different kinematic domains. It was suspected that contradictory conclusion on the values of suppression is due to the different cuts on the $E_{\rm T}$ of the jets in both analyses. This was not fully confirmed, a tendency of weakening of the suppression with $E_{\rm T}$ increasing is observed but within the uncertainties it is still present \cite{Karel,Karel_new}. Here an attempt is made to recalculate (using Monte Carlo model RAPGAP \cite{RAPGAP}) the results from H1 kinematic domain to the one of the ZEUS analysis, see Table 1.

%

Conversion from the H1 phase space to that of ZEUS could be done in two steps.

In the first step the differential cross sections measured by H1 are multiplied in each bin by factors, obtained by MC RAPGAP, accounting on the extension from $Q^2 < 0.01\, \mathrm{GeV^2}$ to $Q^2 < 1\, \mathrm{GeV^2}$. This leads to an increase of the differential cross sections by about a factor of 1.35 with no substantial change of their shapes. We expect that the uncertainty of the  extrapolation in $Q^2$ can be neglected since it is given by quantum electrodynamics only  - in these analyses $E_{\rm T}^2$ is used as a hard scale. Next, the data are divided by a factor of 1.2 to correct the cross section to the elastic case of $M_Y=m_{p}$ of ZEUS.\footnote{The ratio $1.20 \pm 0.11$ is taken from \cite{Misha}.}

In the second step a matrix is constructed, by means of MC RAPGAP, which accounts on transition from the H1 phase space to that one of ZEUS, $ \mathbf{M}_{H\to Z}$. The matrix is determined separately for each pair of corresponding differential cross sections from both analyses ($x_{\gamma}$, $E_T^{\mathrm{jet1}}$,\ldots). The matrix contains  the probabilities that a particular event belonging to the $i$-th bin will end up in the $j$-th one of the given cross section as one makes the transition from the H1 to the ZEUS analysis phase space. 
 Although the shapes of differential cross sections measured by H1 are rather well described by MC \cite{Karel_diss}, to achieve a higher accuracy of the matrix determination, the MC spectra are reweighted to match the data better. 
In addition the contribution of events generated outside the H1 kinematics but fulfilling ZEUS cuts needs to be taken into account. The normalization of this contribution is provided by the previously discussed reweighting of MC. The including of these events depends on the validity of used model in the region uncovered by H1 data.
For each variable the result of the extrapolation procedure is given by a histogram   
 $\mathbf{\sigma}_{Z}$ determined as
\begin{equation}
\mathbf{\sigma}_Z = \mathbf{M}_{H\to Z} \; \mathbf{\sigma}_{H} + \mathbf{\sigma}_Z^{add}
\end{equation}
where $\mathbf{\sigma}_{H}$ is the vector of values of the measured cross sections modified in step one and 
$\mathbf{\sigma}_Z^{add}$ is the contribution from outside of the H1 phase space.

Results of this transformation of H1 data are presented in Fig. 4 together with ZEUS published data. It can be seen that after the transformation to the identical phase space the H1 differential cross sections are  lower than the ZEUS results	 by factor about 0.6. On the other hand the shapes of H1 and ZEUS differential cross sections are very similar,  except for $z_{I\!\!P}$. 
The uncertainties of the extrapolated H1 data were determined by propagation of the statistical and uncorrelated systematic errors from \cite{Karel_new}. The relative uncertainty of  $\mathbf{\sigma}_Z^{add}$ was assumed
to be the same as the one of the total cross section from \cite{Karel_new}. In addition the model dependence of the extrapolation was studied by using two different DPDFs - H1 fit B and H1 fit Jets. This source of uncertainty was found to be relatively small (with the exception of  $z_{I\!\!P}$ distribution) and is included into errors bars shown in Fig. 4.

It is worthwhile to mention that two independent crosschecks of the transformation method were done. The first one is based on the comparison of H1 results from \cite{Karel_new} extrapolated to the kinematic region of H1 preliminary high $E_{\rm T}^{\mathrm{jet}}$
data \cite{Karel}. The extrapolated results agree within errors with cross sections from \cite{Karel} in the shape and normalisation very well. The another crosscheck used ZEUS published data  \cite{Z_dijetsPH}  transformed to the H1 preliminary high 
 $E_{\rm T}^{\mathrm{jet}}$ data \cite{Karel} yielding the same relative difference as shown in Fig. 4. Note that in 
 this case the H1 cuts were the subset of ZEUS cuts and therefore the constant term of the equation (1) was not needed. 

\section{Summary}

The analysis of published H1 and ZEUS results is done with the emphasis to understand better the possible sources of the discrepancies between the conclusions of both collaborations. It is shown that the results are not significantly sensitive to the photon structure function used. Although, here the different hadronisation corrections are obtained than in \cite{Z_dijetsPH}, it has no impact on the interpretation of ZEUS differential cross sections. The conversion of H1 results to ZEUS phase space is done. The shapes of differential cross sections measured by both collaborations are in agreement (except for $z_{I\!\!P}$ variable), however, the H1 results are on average lower by about 40\% than ZEUS ones. Within the limitations of the transformation method (based on Monte Carlo) there is a suggestion that the observed discrepancy between H1 and ZEUS results concerning factorisation breaking is not caused by different phase space of both analyses.

The puzzle of factorisation breaking in diffractive dijet photoproduction could be resolved therefore by new experimental analyses. Among them the most promising one could be the identification of diffractive events based on leading proton detection.  This method has both an advantage of providing a data sample free of proton dissociation and of reducing the uncertainties of the diffractive selection if compared with large rapidity gap method.

\section{Acknowledgements}
We thank to M. Ryskin and V. Khoze for many valuable discussions and reading of the text.
We are grateful to DESY Hamburg for the possibility of using DESY computing resources.
This work was supported by the grant LC527 of MSMT of Czech Republic and grant SVV 263309 of Charles University in Prague.

\end{document}